\documentclass[12pt,a4paper,reqno]{article}

\usepackage{amsmath,latexsym,amssymb,amsthm}
\usepackage{graphicx}
\usepackage{hyperref}
\usepackage{amsmath}\usepackage{amssymb}
\usepackage{amsthm}

\newcommand{\be}{\begin{equation}}
\newcommand{\ee}{\end{equation}}

\renewcommand{\>}{\rangle}

\theoremstyle{definition}

\theoremstyle{remark}

\begin{document}

\title{\bf{The life and entangled adventures of Schr\"odinger's cat}}

\author{A Sudbery$^1$\\[10pt] \small Department of Mathematics,
University of York, \\[-2pt] \small Heslington, York, England YO10 5DD\\
\small $^1$ tony.sudbery@york.ac.uk}

\date{3 August 2023}

\maketitle

\begin{abstract}
A lecture given at St. Cross College, Oxford, on 10th June 2017 in a conference on The Nature of Quantum Reality

In this lecture, intended for a general audience, I describe Schr\"odinger's thought experiment which was designed to show the strange results of extending the formalism of quantum theory, particularly the idea of superposition, beyond the subatomic regime. I describe a way to understand superposition in the terms of formal logic. I trace the development of Schrodinger's ideas after this thought experiment, and briefly sketch some work which realises it in actual experiments, and proposals for taking it further.

\end{abstract}

In this talk I will need to review some of the basic ideas of quantum theory, which means that there will be some overlap with the previous talk by Jim Baggott. And as we all know, in quantum theory where there is overlap there is likely to be interference.

\begin{center}SUPERPOSITION \end{center}

One of the basic ideas of quantum mechanics -- some might say, \emph{the} basic idea -- is \emph{superposition}. This is denoted in the mathematical form of the theory by the simple, innocent symbol $+$. Yet this simple operation takes on a very mysterious aspect when we consider its physical meaning for objects at very small, subatomic scales. Consider a single particle at this scale, and ask ``Where is it?'' We might know that it is at, or near, one definite place, say ``here''. This could correspond to a function of position like $\Psi_1(x)$ in the first slide, a state of affairs which is represented in quantum theory by the symbol $|\text{here}\>$ in a notation invented by Dirac. The characters $|$ and $\>$ indicate that this is a description of the state of a quantum object, and between them is put information about this description. Another possibility is that the particle might be in a different place, say ``there'', corresponding to another function $\Psi_2(x)$ or a state $|\text{there}\>$.

\begin{figure}\label{superposition}
\scalebox{0.5}{\includegraphics{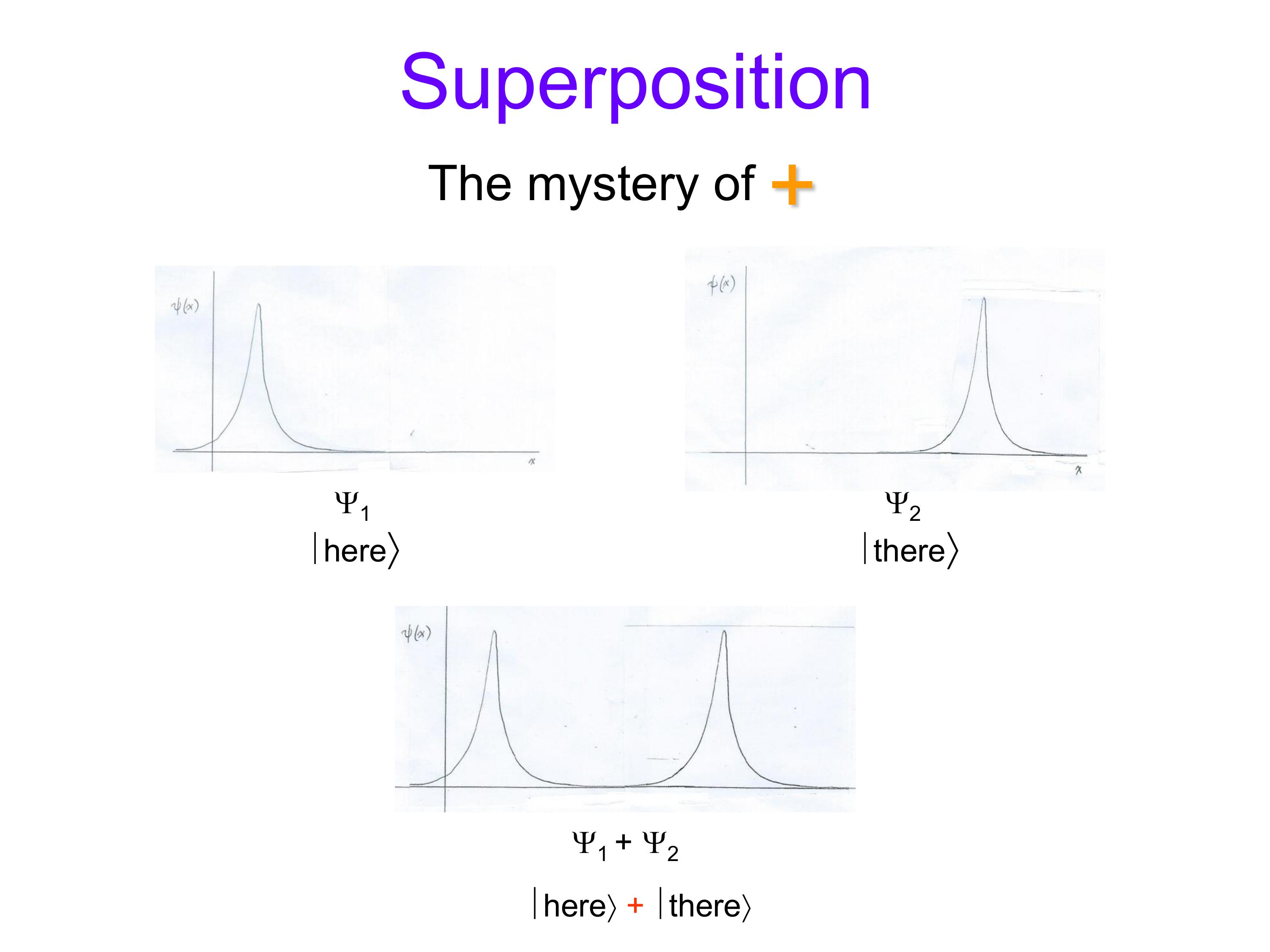}}\hspace{2cm}
\end{figure}

But according to quantum mechanics, there is a third possibility: the particle might be in a \emph{superposition} of being here and being there, corresponding to the function $\Psi_1(x) + \Psi_2(x)$ and represented by the state symbol $|\text{here}\> + |\text{there}\>$. You might think that there is nothing mysterious about this; it simply means that the particle is \emph{either} here \emph{or} there. As long as the functions $\Psi_1(x)$ and $\Psi_2(x)$ do not overlap, this seems reasonable; but if they start to overlap, the phenomenon of \emph{interference} makes itself known, and shows that we cannot understand superposition this way.

 \begin{figure}\label{twoslits}
\scalebox{0.5}{\includegraphics{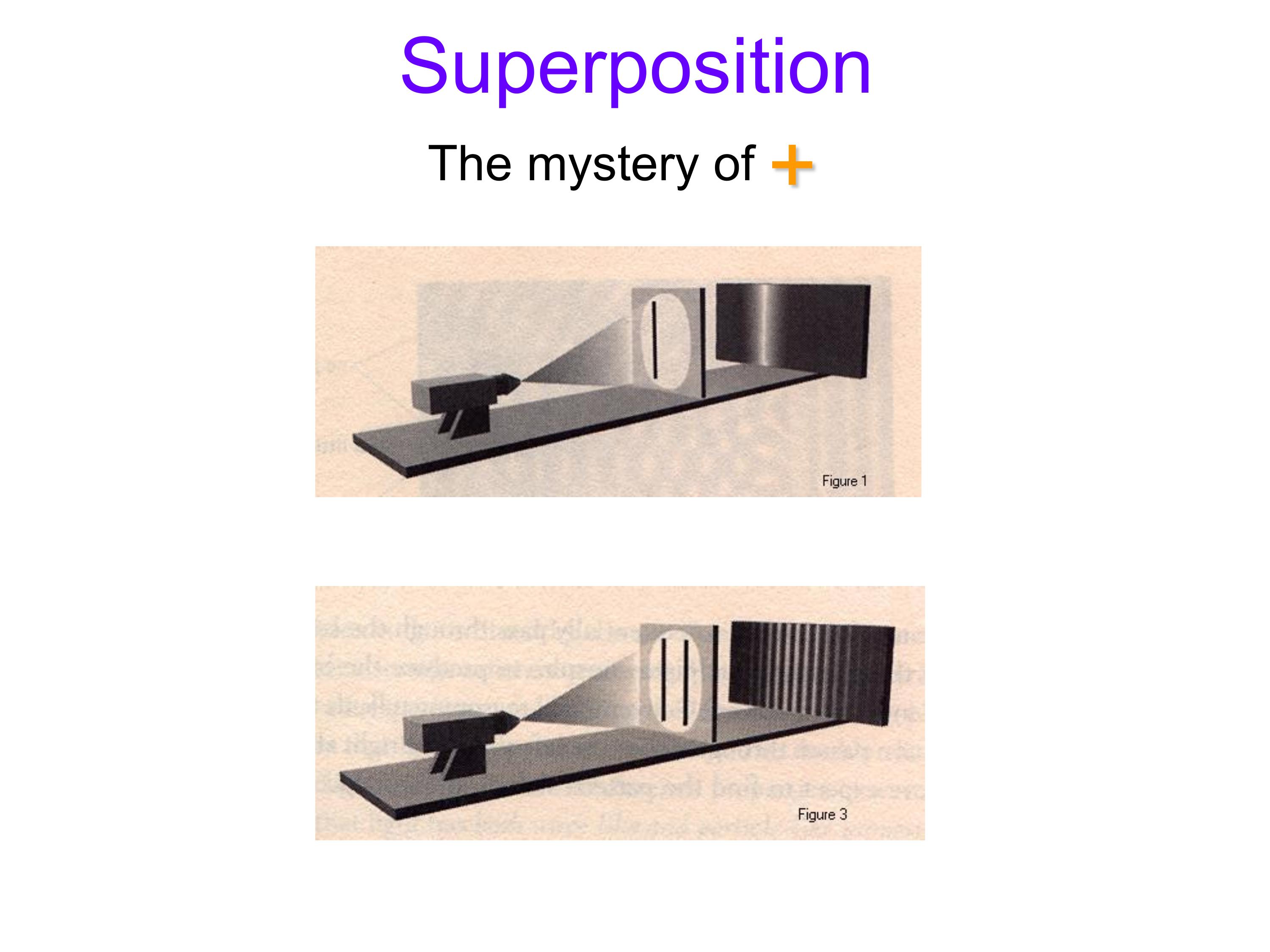}}\hspace{2cm}
\end{figure}

This is demonstrated in the \emph{two-slit experiment}, shown in the next slide. A beam of particles, inaccurately collimated so that they cover a range of directions, is fired at a screen in which two parallel slits have been cut and can be covered by shutters. Beyond this screen is a second screen which collects the particles and registers their arrival. If just one slit is open, the particles reach the second screen in a fuzzy band centred on a line parallel to the line of the slit. What happens if both slits are open? If the particles are the sort of thing we are used to in classical physics, like bullets, then each particle must go through one slit or the other and the pattern registered on the second screen will be two separated fuzzy lines. But that is not what is seen with subatomic particles like electrons. Instead, there is a corrugated pattern, known as an \emph{interference pattern}, as shown in the lower picture of the slide.

The description of this experiment in quantum mechanics is as follows. If only the left-hand slit is open, a particle that goes through that slit will be in a state $|\text{slit 1}\>$ when it reaches the slit, and this will go on to become a state $|\text{line 1}\>$ when it reaches the second slit. There will be other states $|\text{slit 2}\>$ and $|\text{line 2}\>$ describing particles that go through the right-hand slit when that is open. When both slits are open the state of each particle at the second screen will be $|\text{line 1}\> + |\text{line 2}\>$. If the symbol $+$ in that expression meant ``or'' -- if each particle went through \emph{either} the left-hand slit \emph{or} the right-hand slit -- then the pattern on the second screen would be two parallel lines, as it is with bullets. But that is not what is seen. So $+$ does not mean ``or''. It surely doesn't mean ``and'' -- a particle certainly isn't in the left-hand line on the second screen \emph{and} in the right-hand line. That familiar symbol $+$, when applied to descriptions of quantum particles, means something entirely new.

Another new aspect of superposition is that unlike ``and'' and ``or'', which are both single ways of combining descriptions, superposition encompasses a range of possibilities: the two states $|\text{here}\>$ and $|\text{there}\>$ allow superpositions $a|\text{here}\> + b|\text{there}\>$ where $a$ and $b$ are any numbers (complex numbers, even) though it is only their ratio $a/b$ that counts. Thus superpositions of $|\text{here}\>$ and $|\text{there}\>$ range from the simple $|\text{here}\>$ continuously through the equal superposition $|\text{here}\> + |\text{there}\>$ to the simple $|\text{there}\>$ -- something which has to be taken into account in any explanation of the quantum meaning of $+$.

\begin{center} ENTER THE CAT \end{center}

In 1935 Erwin Schr\"odinger faced up to the implications of this new theory for what physics has to tell us about the whole world. If quantum mechanics is the basic theory of how matter behaves, then it should apply to all matter, including familiar objects like animals. This strange new concept $+$ should apply to such familiar things; if something can exist in a state $|\text{here}\>$, and it can also exist in a different state $|\text{there}\>$, then this physics says that it can also exist in a third state $|\text{here}\> + |\text{there}\>$. He wrote:

\begin{quote} One can even set up quite ridiculous cases. A cat is penned up in a steel chamber, along with the following diabolical device (which must be secured against direct interference by the cat); in a Geiger counter there is a tiny bit of radioactive substance, \emph{so} small, that \emph{perhaps} in the course of one hour one of the atoms decays, but also, with equal probability, perhaps none; if it happens, the counter tube discharges and through a relay releases a hammer which shatters a small flask of hydrocyanic acid. If one has left this entire system to itself for an hour,, one would say that the cat still lives \emph{if} meanwhile no atom has decayed. The first atomic decay would have poisoned it. The $\psi$-function of the entire system would express this by having in it the living and the dead cat (pardon the expression) mixed or smeared out in equal parts. \cite{Schrcat} \end{quote} 

In the Dirac notation introduced above: if a cat can exist in a state $|\text{alive}\>$ and can exist in a state $|\text{dead}\>$, as it surely can, then it can also exist in a state $|\text{alive}\> + |\text{dead}\>$.

\newpage

\begin{center} AND YOUR POINT IS \ldots ? \end{center}

The wording at the end of that passage by Schr\"odinger is a little misleading. Elsewhere in the same article Schr\"odinger was at pains to insist that superposition is \emph{not} a matter of ``smearing'':

\begin{quote} There is a difference between a shaky or out-of-focus photograph and a snapshot of clouds and fog banks. \end{quote}

But whatever superposition is, Schr\"odinger regards this consequence of orthodox quantum mechanics as ``ridiculous''.

This point is often made (but \emph{not} by Schr\"odinger) by arguing ``If cats can exist in this strange state, why don't we ever see such (alive + dead) cats? The way to answer such questions was indicated by Schr\"odinger himself; it lies in the concept of \emph{entanglement} which he introduced in this same paper.

\begin{figure}\label{why no see}
\scalebox{0.5}{\includegraphics{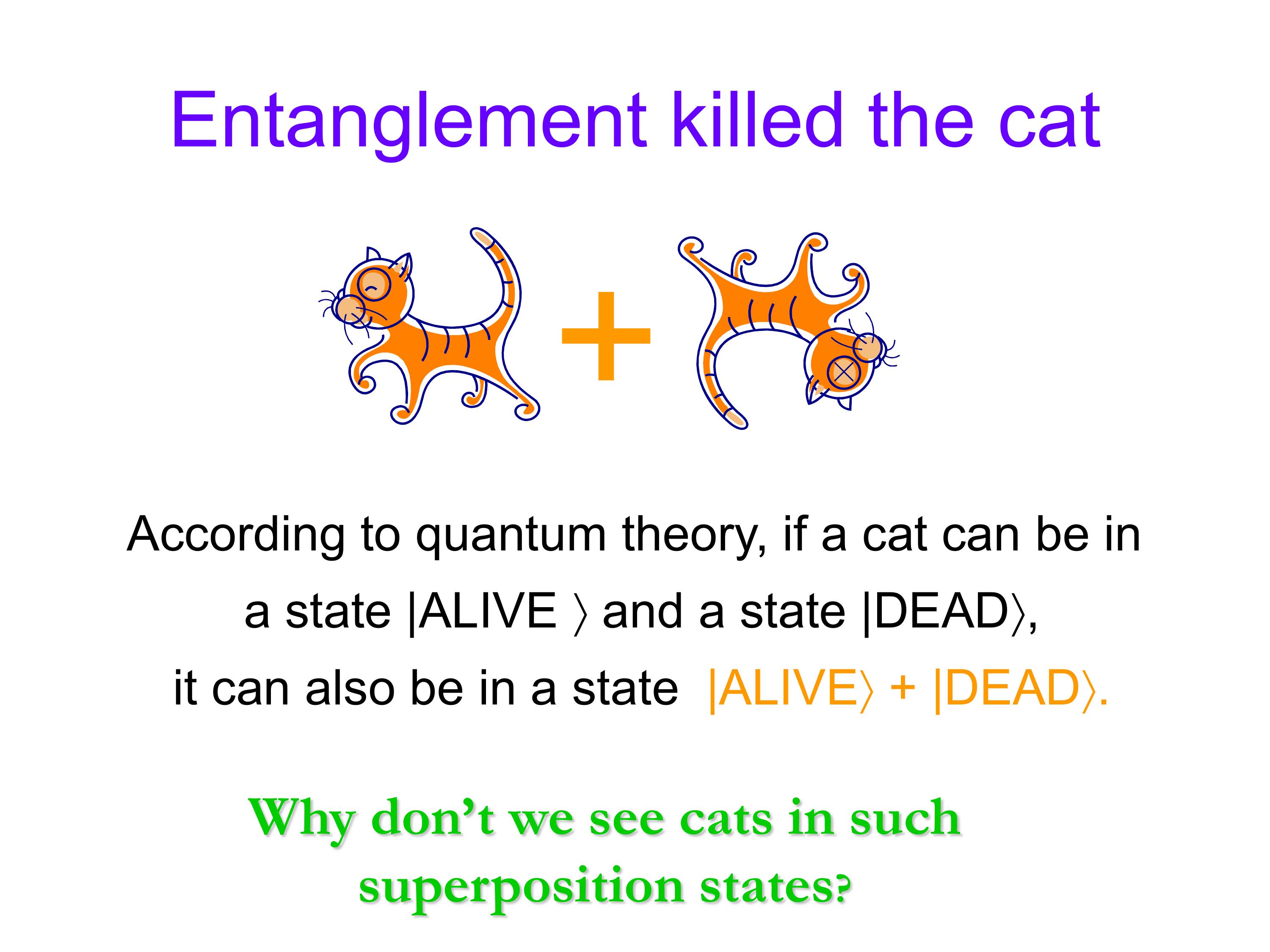}}\hspace{2cm}
\end{figure}

\begin{center} ENTANGLEMENT \end{center}

\begin{figure}\label{because}
\scalebox{0.5}{\includegraphics{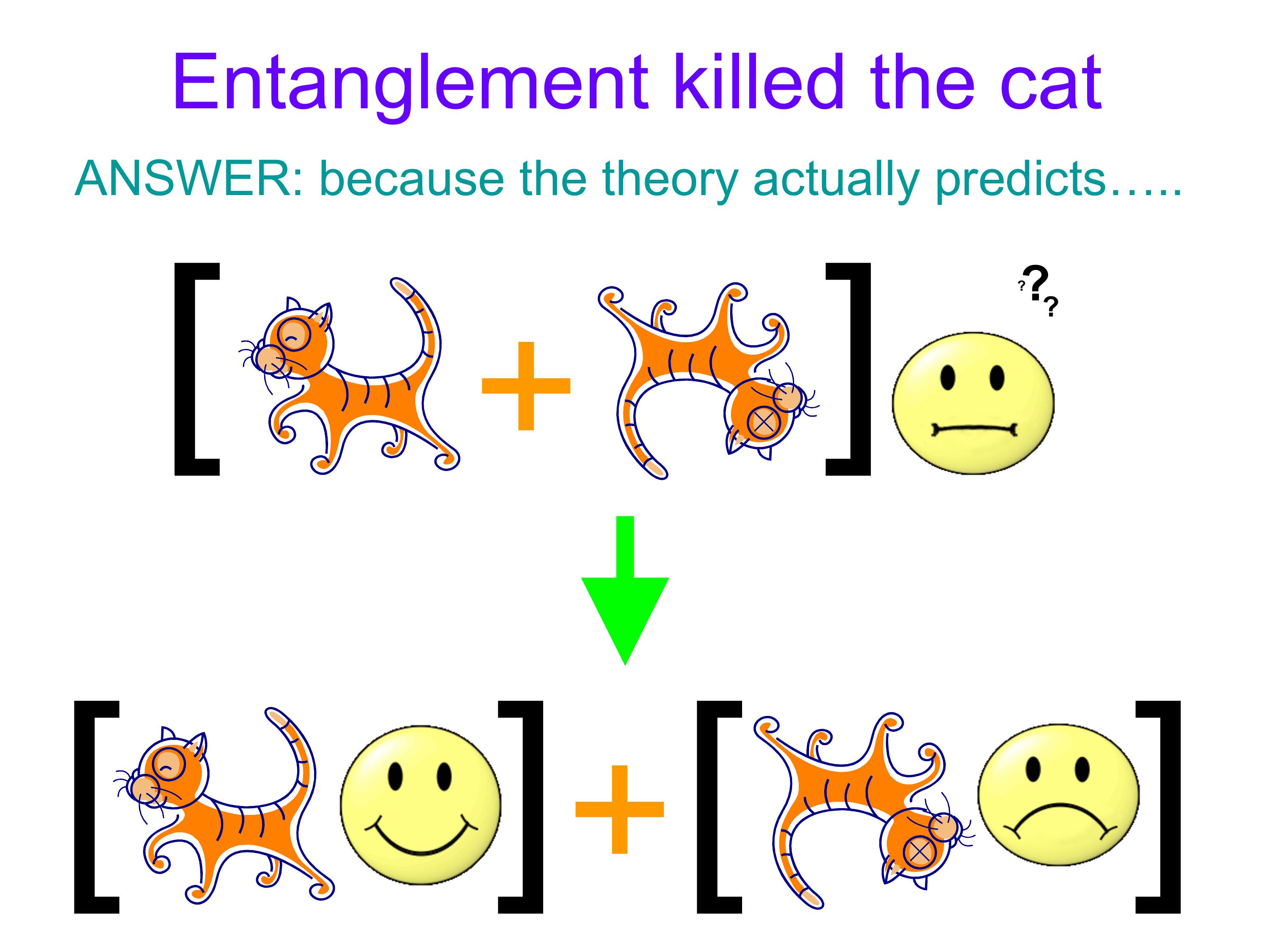}}\hspace{2cm}
\end{figure}

The answer to the question ``Why don't we see cats in superposition states like $|\text{alive}\> + |\text{dead}\>$?'' is that the physics of looking at a cat shows that the observer will never see such a thing. When the box containing the cat in Schr\"odinger's diabolical device is opened, the cat is not in the state $|\text{alive}\> + |\text{dead}\>$; instead, the whole system of the cat and the observer (and the rest of the device) is in the superposition state
\be\label{entangled}
|\text{alive}\>|\text{glad}\> + |\text{dead}\>|\text{sad}\>.
\ee
Here the first term is a state of the cat and the observer in which the cat is alive and the observer is glad because they see that the cat is alive; likewise, the second term is a state of the cat and the observer in which the cat is dead and the observer is sad because they see that the cat is dead. Nowhere is there a state in which the observer sees the cat in a superposition. This is the case because we know that the ordinary processes of everyday physics will cause the cat state $|\text{alive}\>$, together with an observer outside the closed box, to become the first term in eq. \eqref{entangled} when the box is opened; likewise, the cat state $|\text{dead}\>$ with the observer outside the box will become the second term. The rules of quantum mechanics then guarantee that the superposition state of the cat, with the observer outside the box, will become the combined state \eqref{entangled} when the box is opened. 

The state \eqref{entangled} is said to be an \emph{entangled} state of the combined system of the cat and the observer. Neither the cat nor the observer has a definite state, but they are correlated: if the cat is (somehow) found to be alive, then the observer will be found to be glad, and if the cat is found to be dead, then the observer will be found to be sad. The word ``entanglement'' was introduced by Schr\"odinger (translating the German Verschr\"ankung) in the paper \cite{Schrcat} that introduced the cat; the concept was already implicit in the paper by Einstein, Podolsky and Rosen which prompted Schr\"odinger's paper -- it is just a special case of superposition --  but Schr\"odinger recognised its special significance, and it has continued to play a central role in discussions and applications of quantum mechanics.

\begin{center}MEASUREMENT\end{center}

I hear impatient mutterings in the audience: ``What do you mean, if the cat is (\emph{somehow}) found to be alive. If someone opens the box and sees the cat alive, then surely the cat \emph{is} alive.'' Well, you would think so, wouldn't you? But the rigorous theory of quantum mechanics, in the form of the equation of motion (due to -- guess who -- Schr\"odinger) which governs the change of the state vector, only gives the superposition shown in eq. \eqref{entangled}. To the founders of quantum mechanics, who regarded the results of observation, or \emph{measurement}, as the only facts that belong in science, this was not a problem: one calculates using the Schr\"odinger equation and then uses a well-defined procedure (known as the \emph{Born rule}) to obtain probabilities for the possible results of measurement. One can then throw away the state vector as no longer meaningful. Or if one wants to make further calculations, one has to allow for an unpredictable change caused by the very fact of measurement, which cannot be further analysed. This unpredictable change is described by a supplement to the equation of motion known as the \emph{projection postulate} (due to Paul Dirac, though it is often erroneously credited to John von Neumann). 

The resulting complete form of quantum theory, which is presented in most textbooks, presents two different laws for the change of physical objects, or systems of objects: the Schr\"odinger equation and the projection postulate. The first of these applies as long as the objects are not disturbed by a measurement. This is deterministic: given the state of the objects at one time, the state at any later time is uniquely determined by the equation. It's a beautiful law of motion for any physical system. But, according to official quantum mechanics, this motion is rudely and randomly interrupted whenever someone makes a measurement. Only probabilities can be given for the result of the measurement, and the system changes, depending on the result of the measurement, according to the projection postulate. 

This situation was acceptable to the founding fathers, who were struggling to understand the strangeness of the subatomic world which was only just being revealed. But increasingly, modern physicists find it unacceptable. In the case of Schr\"odinger's cat, does it mean that the RSPCA official who rushes into Schr\"odinger's lab to stop this diabolical experiment actually kills the cat, or possibly saves it, simply by opening the box to see if the cat is alive? More seriously, what is this ``measurement'' that causes the system to behave in a completely different way? Any measurement or observation is performed by an arrangement of physical apparatus, which should be subject to the Schr\"odinger equation like any other physical objects. What makes it behave differently? This is the \emph{measurement problem} of quantum mechanics, which still has no generally accepted solution. It is often raised by students, who may be met with impatient dismissal and told, as David Mermin put it, to ``shut up and calculate''. I was a student in a class given by Dirac, who, when we raised this problem, simply shrugged and said ``That's how it is''.

\begin{center}THE HISTORY OF SCHR\"ODINGER'S CAT \end{center}

There's a fork in the path here. First, let's follow Schr\"odinger.

\begin{center}THE HISTORY OF SCHR\"ODINGER'S THOUGHT \end{center}

In 1926 Schr\"odinger didn't \emph{invent} quantum mechanics, but he \emph{explained} quantum mechanics as an eigenvalue problem in terms of a wave function, which is the sort of thing that physicists like to deal with. And everybody -- no, not everybody, but a lot of practical people using quantum mechanics, and in particular chemists -- are really happy working with wave functions rather than the alternative mathematical description of Heisenberg and Dirac in terms of state vectors, which is often seen as "abstract'' (though I think that on the contrary, Dirac's formalism is more concrete). Anyway, Schr\"odinger was attached to the wave function, and to the wave equation. In 1935 he presented the story of the cat as an illustration. (It's interesting that he talks about the wave function \emph{of the cat}. That's an enormously complicated function.) In 1952 \cite{Schrodinger:interpretns} he emphasised the importance of the wave function, and in particular the wave equation. He said there are no quantum jumps \footnote{Schr\"odinger had always thought this: in 1929 he grumbled ``If I had known we were going to go on having all this damned quantum-jumping, I would never have got involved in the subject'' \cite{Schrquote}}, nothing changes discontinuously; he even said there are no particles. He didn't believe in particles, he only believed in the wave function. And he made the point (he also made this point in the cat paper) that if you've got a theory of waves, OK you can talk about the waves and the wave front and so on, but a theory of waves also encompasses something very particle-like. In most situations, if you've got a wave developing, then you can interpret this in terms of \emph{rays}. You can do geometrical optics as well as wave optics; and those rays can be interpreted as the trajectories of particles. For Schr\"odinger, a wave description is better, more encompassing, than a particle description. He thought people concentrate far too much on particles. He prefaced this paper by saying ``People are going to think I'm mad. I'm going to tell you that both of these things are true. You've got a particle here \emph{and} the same particle there. You've got a cat dead \emph{and} a cat alive.'' Well, that's a bit mysterious, and in 1955 he cleared it up by saying ``I really do believe that the wave function (of the universe) is all that there is'' \cite{Schrodinger:interpretns}.  This is getting very close to what, much more famously, was published by Everett and endorsed by Wheeler in 1957, and which, after being popularised by de Witt in 1971, has come to be known as the ``many worlds interpretation'' of quantum mechanics. According to this, what that superposition of a live cat (and a happy observer) and a dead cat (and a sad observer) really means is that there are two worlds: one in which the cat is alive and one in which the cat is dead. So Schr\"odinger anticipated the many-worlds interpretation.

This takes us back to the question of the meaning of superposition. What does $|\text{here}\> + |\text{there}\>$ mean? Does it mean that either the particle is here \emph{or} the particle is there? That's what it means for us when we observe the particle; that's the way we tend to think of superposition. But as we saw earlier in the two-slit experiment, it cannot be the true meaning of superposition. It seems clear that it cannot mean that the particle is here \emph{and} the particle is there, though the many-worlds interpretation seems to be a way of trying to make sense of this. But there are problems with this; who decides what exactly a ``world'' is? I don't think there are many worlds; there is only one world, and its state is a superposition. I would like to show you a way to make sense of this.

\begin{center}LOGIC\end{center}

There is a lot of discussion about the ``reality'' of the wave function. But that's a category error. We're not talking about things -- whether the wave function is \emph{real} -- we're talking about propositions: whether the wave function is a \emph{true} description. If we have two propositions $P$ and $Q$, then the compound proposition $P\;or\;Q$ means that just one of $P$ and $Q$ is true, i.e. has truth value 1. The compound proposition $P\;and\;Q$ means that they both have truth value 1. In quantum mechanics, with $P$ and $Q$ being state descriptions (or state vectors) we have this new way, superposition, of forming compound propositions: $aP + bQ$ where $a$ and $b$ are complex numbers which we can take to satisfy $|a|^2 + |b|^2 = 1$. Now the truth value is not just 0 or 1 for each of these propositions, but the truth is divided between them. We have to use a many-valued logic in which truth values, just like probabilities, can take any value, a real number beteen 0 and 1. So if the universe is described by a superposition of states, in one of which the cat is alive and in the other it is dead, then it's that superposition that has truth value $1$, and the individual components -- ``cat alive'' and ``cat dead'' -- are not true, at least they're not completely true; but they're not completely false either. If $a|\text{alive}\> + b|\text{dead}\>$ is true, then $|\text{alive}\>$ has truth value $|a|^2$ and $|\text{dead}\>$ has truth value $|b|^2$. 


There's another logical point. I've been talking about the state of the whole universe; that's a sort of God's eye view, which we can never attain -- still, we might believe that it exists. But that's not what we actually have. We are inside the universe. Living in this universe with the cat in it, in the superposition shown earlier, I am one of these people -- that is, I have one of these brain states. Another way to understand this superposition is to say that it's a conditional statement. It says that if I know that one of these partial states (of part of the universe) is true, however I know it, then I know that the state of the rest of universe which is attached to it is also true. If I know that I am happy, then this superposition state tells me that the cat is alive. And of course, I do know that I am happy, or sad, or puzzled. But this contradicts the quantum-mechanical view, the God's-eye view, of the universe. So we have to acknowledge context, or perspective. From my perspective, I know that I am happy, or sad, or not knowing; and I know these statements with the usual bivalent logic. There is only a certain set of experiences that I can have (determined by the physics of my brain), and I know that just one of them is true of me at the present moment. Only one of them has the truth value 1; the others have truth value 0, and there is nothing in between. 

Ordinary bivalent logic applies to statements that I make about my present experience. But it is different for statements about my future. According to the Schr\"odinger equation, my present state of a definite experience can evolve into a superposition of different experience states. In that situation, what can I say about what I will experience at some point in the future? My state will be a superposition; but I can't experience such a state. I will have some definite experience, but there is no true statement about which experience it will be. Each statement ``I will have experience $E$'' has a truth value between $0$ and $1$; or, as we usually say, there is only a probability that I will have that experience. That probability, or truth value, is given by the squared modulus of the coefficient of $|E\>$ in my future superposition state.

This is not a new idea; it goes back to Aristotle. In a famous passage, Aristotle discussed the truth or falsity of the proposition ``There will be a sea battle tomorrow''. He concluded that such a statement was neither true or false. This has been taken by modern logicians to mean that statements in the future tense obey a many-valued logic; they have augmented the usual values $1$ or$t$ (true) and $0$ or $f$ (false) by a third truth value $u$ (undecided). For formal reasons, they have restricted truth values to these three, but it seems much more fitting in a temporal logic \cite{logicfuture} to use the full set of values between $0$ and $1$, which relates them to probability. Aristotle, indeed, pointed out that future events can be more or less likely. 

\begin{center}THE HISTORY OF SCHR\"ODINGER'S CAT\end{center} 

Let's get back to the adventures of the cat. It was 82 years ago that Schr\"odinger told the story of the cat. In the first 40 years after that, only 26 people mentioned it in a scientific paper. That might reflect citing habits, because by 1975, I can assure you, the idea of Schr\"odinger's cat was well known. But in 1986 there was a symposium \cite{Schrodinger:centenary} to celebrate the centenary of his birth. There were 19 papers, and only one, by John Bell, gave a citation to the cat paper. The word ``cat'' does not appear in the index. People who were interested in Schr\"odinger knew about his cat -- John Bell's mention of it makes it clear that he expects people to know what he's talking about -- but it wasn't regarded as something of serious scientific interest. But since then there have been over 300 citations of the original paper in German; I didn't count references to the English translation, which appeared in 1980 and was reprinted in an influential collection of papers edited by Wheeler and Zurek \cite{QTMeasurement}. Never mind citations: last year alone there were 119 papers with ``Schr\"odinger's cat'' in the title. And these are not fuzzy philosophy: about two-thirds of these papers are hard physics, concerned with actual experiments.

I've said that the standard idea of Schr\"odinger's cat, i.e. a cat which is, on its own, in a superposition of alive and dead, is nullified by the fact that this state cannot, in practice, appear on its own, because it is continually being entangled with its environment: with molecules in the atmosphere, with electromagnetic radiation, and so on. But that state of an isolated system is what has come to be known as a ``Schr\"odinger cat state'': an actual state of a macroscopic object, separated from the rest of the universe, isolated from any interaction with anything else, and in a superposition of two states which we would recognise as being different at a macroscopic level. So there is a challenge: can experimenters actually make objects which really are in a superposition in this sense?



\newpage

\begin{center}DISENTANGLED ADVENTURES\end{center}

Nobody has done it with a cat (yet), but there is talk of doing it with a living object. In 1996 Wineland's group constructed a superposition state of a single atom \cite{Wineland:Be}, demonstrated by the fact that a beam of these atoms, put through two slits, did show interference as in my second slide. Three years later Zeilinger's group did it with a large molecule \cite{Zeilinger:C60}. If you can do it with a large molecule, why not a virus \cite{Cirac:virus} -- a virus is alive (sort of) -- and if with a virus, why not a microbe?


\begin{figure}\label{MontyPython}
\scalebox{0.6}{\includegraphics{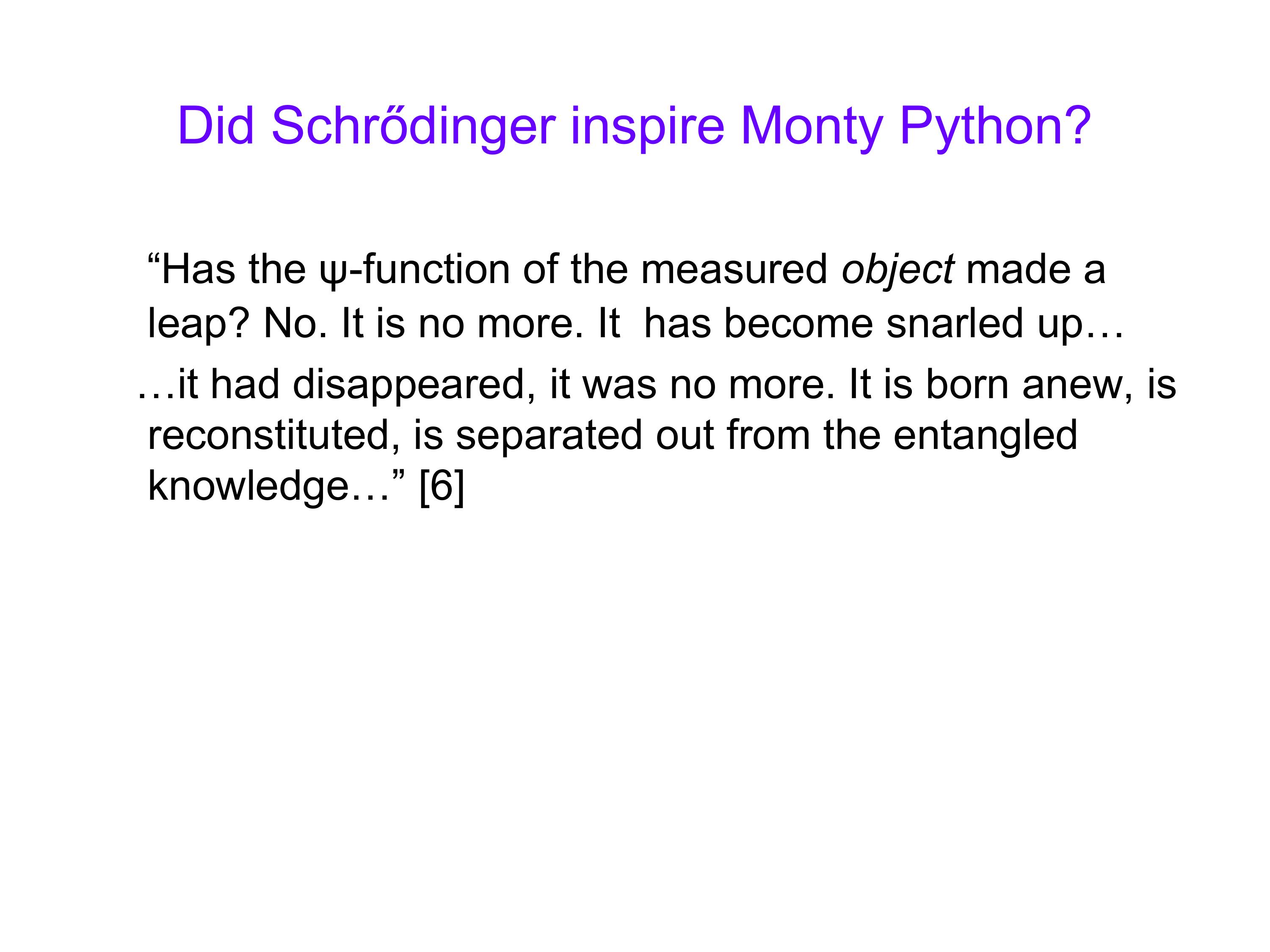}}\hspace{2cm}
\end{figure}

\end{document}